# Compact Device Models for FinFET and Beyond

Darsen D. Lu[1], Mohan V. Dunga[2], Ali M. Niknejad[3], Chenming Hu[3], Fu-Xiang Liang[1], Wei-Chen Hung[1], Jia-Wei Lee[1], Chun-Hsiang Hsu[1], Meng-Hsueh Chiang[1]

[1]National Cheng Kung University, Tainan, Taiwan R.O.C.
[2]SanDisk Corporation, Milpitas, CA, USA
[3]University of California, Berkeley, CA, USA
E-mail: darsenlu@mail.ncku.edu.tw

**Abstract** Compact device models play a significant role in connecting device technology and circuit design. BSIM-CMG and BSIM-IMG are industry standard compact models suited for the FinFET and UTBB technologies, respectively. Its surface potential based modeling framework and symmetry preserving properties make them suitable for both analog/RF and digital design. In the era of artificial intelligence / deep learning, compact models further enhanced our ability to explore RRAM and other NVM-based neuromorphic circuits. We have demonstrated simulation of RRAM neuromorphic circuits with Verilog-A based compact model at NCKU. Further abstraction with macromodels is performed to enable larger scale machine learning simulation.

## 1. Introduction

The concept of compact device models exist since SPICE [1] was introduced to the world about than 40 years ago. Compact device models, or compact models, consist of a set of analytical expressions that can be as simple as Ohm's Law for the current-voltage relationship of an ideal resistor, or a significantly larger set of expressions and more than 100 model parameters for the case of state-of-the-art MOSFET compact models such as BSIM4 [2]. The evolvement of compact model development is strongly tied to device technology. The FinFET technology has been introduced in 1999 [3] and started to become mainstream CMOS about a decade later [4]. BSIM-CMG [5], a compact model for common multi-gate MOSFET transistor, was introduced and selected as industry standard FinFET model in 2012 in anticipation for the technology change. With the rapid growth of semiconductor technology, a wide range of new electronic device have emerged, such as tunneling FET, negative capacitance FET for CMOS, resistive memory (RRAM), phase change memory (PCM) for non-volatile memory, and organic thin-film transistors (OTFT) for display technology and RFID's. The Verilog-A language [6] has facilitated the quick development of compact models for these new devices, such as those published on the NEEDS online platform [7].

The availability of compact models for many new devices facilitates the evaulation of new circuit applications for emerging technologies. One such application is the design of neuromorphic circuits and systems with emerging non-volatile memory devices, such as RRAM, PCM, spin-torque transfer magnetic memory (STT-MRAM) and ferroelectric memory (FeRAM). We may use emerging memory compact models to perform neuromorphic simulation in the circuit and system level. This allows us to link fundamental device characteristics to the performance of artifical neural network implemented using neuromorphic circuits – a proimsing future direction for artificial intelligence (AI) systems.

In this paper, we first introduce BSIM-CMG and BSIM-IMG, focusing on their surface-potential-based model formulation and versatile modeling capability to cover many different multiple-gate devices. Subsequently, we discuss compact models for emerging memory devices, using RRAM compact model as an example [8]. We have successfully applied it to simulate simple neuromorphic systems. On-going work is to build a neuromorphic simulation platform that links the characteristics of any novel memory devices to the energy consumption, computational speed and throughput, as well as classification accuracy of future AI systems based on novel memory devices.

## 2. Multiple-Gate MOSFET Compact Modeling

As CMOS scaling has reached sub-50nm technology nodes, control of short channel effects becomes extremely difficult. The introduction of High-K and metal gates allowed continued scaling for about two more generations [9]. Further scaling had required multiple-gate MOSFET, which included thin-BOX FD-SOI [10] (Fig. 1(a)) and FinFET (Fig. 1(c)(d)) devices. The former is modeled by BSIM-IMG, where the double-gate structure is considered and the front- and back-gate stacks are allowed to be asymmetric; the latter is modeled by BSIM-CMG, where the two sides of the double-gate structure is assumed to be symmetric. Even though BSIM-IMG is originally designed for planar FD-SOI with back-gate, it can also be used to model independent-gate FinFETs (Fig. 1(b)). BSIM-CMG may also be used to model nanowire MOSFETs (Fig. 1(e)) and vertical transistors (Fig. 1(f)).

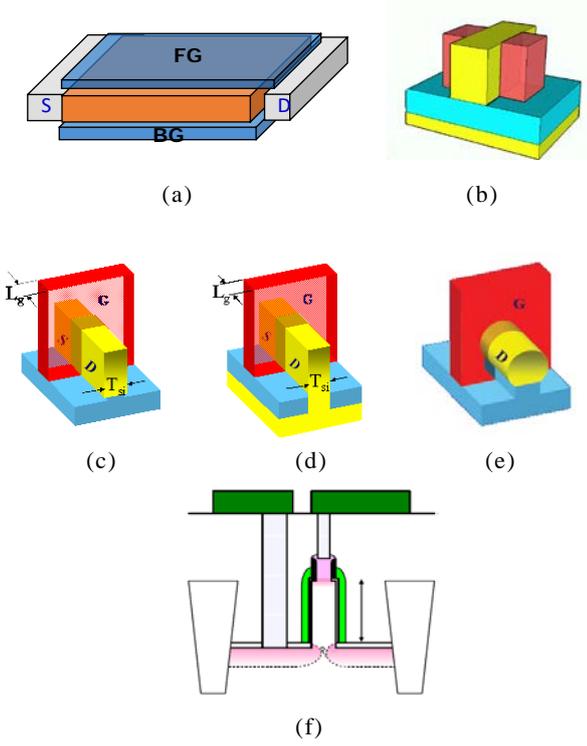

(a)

(b)

(c)　(d)　(e)

(f)

Fig. 1 Multiple-gate Device Types covered by BSIM-CMG and BSIM-IMG. (a) Back-gated UTB-SOI MOSFET and (b) independent-gate FinFET devices are modeled by BSIM-IMG, whereas (c) SOI and (d) bulk FinFET, (e) nanowire FET and vertical pillar FET's are modeled by BSIM-CMG.

Both BSIM-IMG and BSIM-CMG are modeled using surface-potential-based formulations. For example, BSIM-CMG starts from the very basic Poisson's equation considering inversion electrons and body doping:

$$\frac{\partial^2 \psi(x,y)}{\partial x^2} = \frac{qn_i}{\varepsilon_{Si}} \cdot e^{\frac{q(\psi(x,y)-\phi_B-V_{ch}(y))}{kT}} + \frac{qN_A}{\varepsilon_{Si}} \quad (1)$$

where $V_{ch}$ is the channel potential, which varies from 0V at the source end to drain-to-source voltage $V_{ds}$ at the drain end, and $N_A$ is the body doping concentration. Even with several simplifying approximations, the solution to (1) for a given $V_{ch}$ is still an implicit expression which takes the following form:

$$f(\psi_s) = 0 \quad (2)$$

where $\psi_s$ is the surface potential, the potential at the surface of the MOSFET. Third order Householder's method is applied to express $\psi_s$ as an explicit function of boundary conditions such as the gate voltage $V_g$. Such approximation has worked very well for most $N_A$ values and geometry such as body thickness ($T_{si}$) except for extreme values of $N_A$ and geometry. Subsequently, drain current ($I_d$) and terminal charge ($Q_s$, $Q_d$, and $Q_g$) are derived as explicit function of $\psi_{ss}$ ($\psi_s$ at the transistor source terminal) and $\psi_{sd}$ ($\psi_s$ at the transistor drain terminal). The complete expression for $I_d$ is as follows:

$$I_d = \mu \frac{W}{L} \left[ \frac{Q_i^2}{2C_{ox}} + 2V_t Q_i - V_t (5C_{Si}V_t + Q_B) \ln(5V_t C_{Si} + Q_B + Q_i) \right]_s^d \quad (3)$$

where

$$Q_{is(d)} = C_{ox}(V_g - V_{fb} - \psi_{ss(d)}) - Q_B \quad (4)$$

$$Q_B = qN_A \frac{T_{si}}{2} \quad (5)$$

For details of the derivations for $I_d$ and complete C-V expressions, one may refer to [5].

In addition to the aforementioned core (long-channel) $I_d$ and charge models, BSIM-CMG consists of a number of models for real device effects, such as field-dependent mobility, velocity saturation, source velocity limit, parasitic resistances and capacitances, leakage currents, etc. The overall model is evaluated against BSIM4 in terms of computational efficiency (Fig. 2). It was shown that the surface-potential-based BSIM-CMG has similar speed compared to $V_{th}$-based BSIM4 model, which has been widely applied to real-world circuit design.

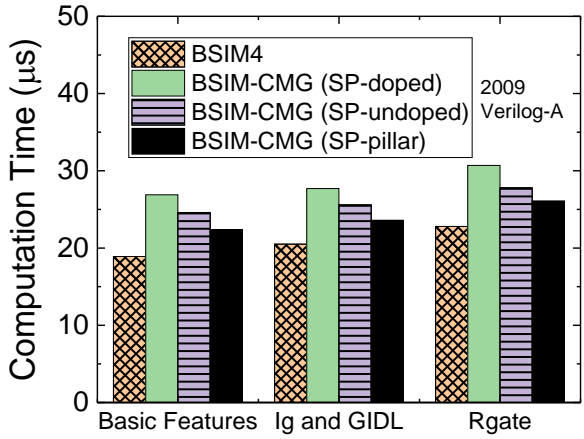

Fig. 2 Evaluation results for computational efficiency of a variety of Verilog-A models (BSIM4 and BSIM-CMG) and surface potential calculation options. "SP-doped," "SP-undoped" and "SP-pillar" refer to the standard surface potential (SP) calculation considering doped body, simplified SP calculation assuming lightly-doped body, and SP calculation for cylindrical geometry, respectively [5].

The derivation for BSIM-IMG follow similar surface-potential-based approach. However, the asymmetric front- and back-gate stacks complicates the calculation of surface potential, $I_d$ and charge. Details of model derivation is available in [11]. BSIM-IMG has been validated against various thin-BOX fully-depleted SOI technology with excellent agreements, e.g. [12].

## 3. RRAM Compact Modeling

Compact modeling techniques can also be applied to memory devices. In Verilog-A and the SPICE environment, memorization of the system state is done by either:
1. Storing information in a real-valued variable in Verilog-A.
2. Representing the system state using the voltage of an extra node.

Typically, the latter is preferred since the life cycle of Verilog-A variables is not well-defined across all simulation platforms. In fact, in BSIM models, history-dependent phenomenon such as self-heating and floating body effect are modeled by introducing the temperature node and the floating-body node, respectively [13]. Similarly, for modeling RRAM characteristics, an R-C network is introduced to model the time-dependent

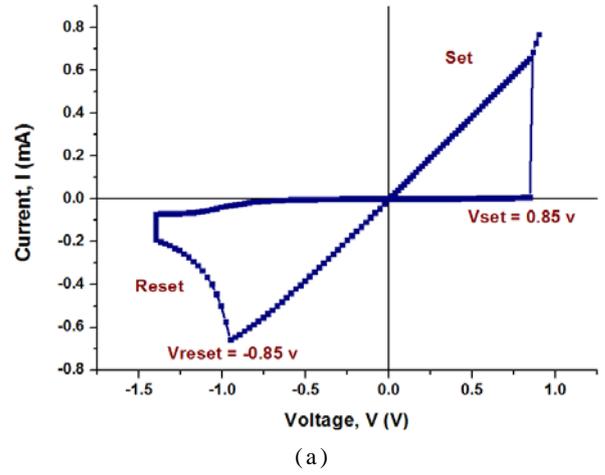

(a)

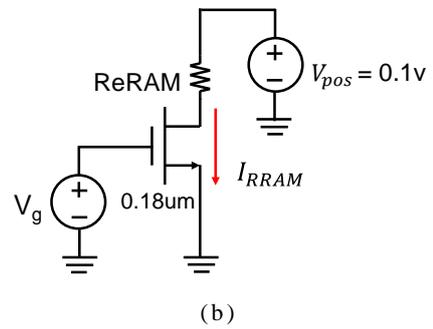

(b)

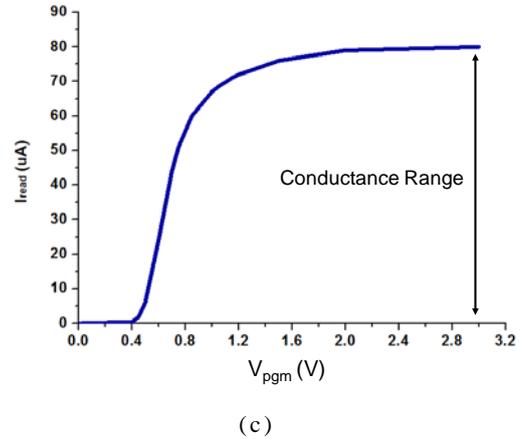

(c)

Fig. 3 Simulation results for Verilog-A based RRAM compact model [8]: (a) SET-RESET I-V characteristics (b) Simulation setup for the 1T-1R circuit (c) Illustration of continuous conductance tuning with various SET compliance current, or MOS gate voltage during SET operation. (Vpos=1.0V for SET operation)

RESET process [8]. Fig. 3(a) illustrates the typical bipolar-switching RRAM I-V characteristics as generated using transient SPICE simulation. The model parameters were calibrated to an $HfO_2$-based RRAM technology [8]. Fig. 3(b) illustrates the 1-transistor-1-RRAM (1T1R) circuit block, which is often used in memory arrays for better isolation. This is as opposed to the cross-bar array, which can cause unwanted leakage across unselected cells. Fig. 3(c) illustrates the continuous resistance tuning simulation. The RRAM is initially at the RESET state with high resistance. $V_{pgm}$ is the voltage applied to the transistor gate terminal for programming (or SET operation, $V_{pos}$=1.0V). $I_{read}$ is the resulting RRAM current during read ($V_{pos}$=0.1V). The fact that RRAM conductance can be continuously varied over a wide range makes it attractive as candidate for neuromorphic circuit design, wherein the RRAM conductance represents the synaptic weights in neural networks.

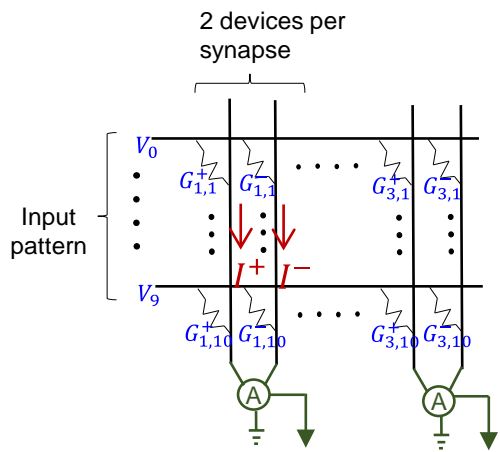

Fig. 4 RRAM-based neuromorphic array. Each synapse of the neural network consists of two RRAM device, whose conductance difference represents synaptic weight [15]. The input pattern (voltage value or pulse train) is passed on word lines; the output current is taken at bit lines, with amplification circuitry to reading out information and providing virtual ground.

## 4. Neuromorphic Circuit Design with RRAM

The compact model for RRAM is applied to the design of neuromorphic circuits [14]. Fig. 4 shows the array configuration considered in this study. Since the conductance of RRAM is always positive, we take the conductance difference between two RRAM devices as the weight of the synapse in question [15]. The input pattern (fixed voltage or pulse train with varying numbers and durations) is passed on horizontal (word) lines; the output current is taken at vertical (bit) lines, with amplification circuitry at each bottom end to read out information and providing virtual ground. We have successfully applied this circuit configuration and offline backward-propagation supervised learning algorithm to train a 3x1 neural network to reproduce the "AND" logic operation, as demonstrated in SPICE [14]. The activation function is implemented offline with software.

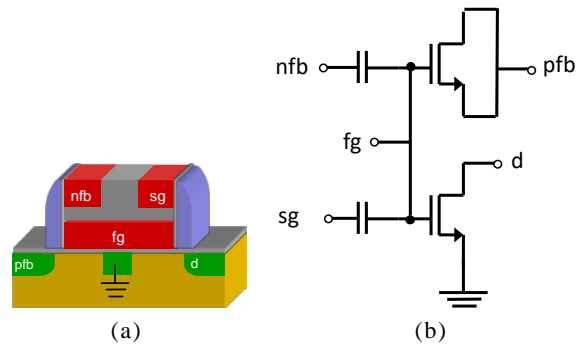

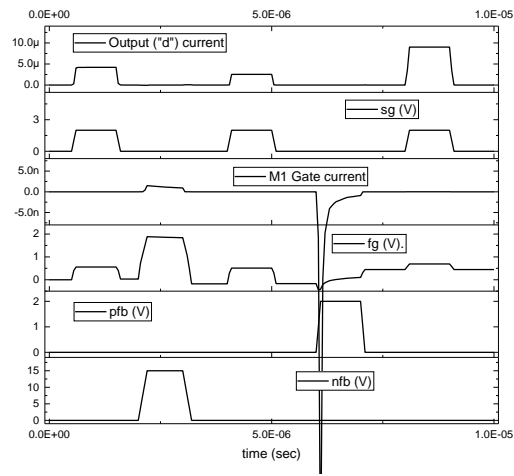

Fig. 5 Simulation of a novel floating-gate synaptic transistor. (a) Device structure with separate negative feedback gate (nfb) for programming and synaptic gate (sg) readout. (b) Equivalent circuit diagram for compact modeling (c) Simulation results with illustrated weight (drain current) weakening after negative feedback pulse and weight strengthening after positive feedback (pfb) pulse.

## 5. Floating-gate Based Neuromorphic Circuits and System Level Simulations

Neuromorphic circuit may also be designed with different types of memory devices. Essentially, this allows us to avoid separation of computation (such as the CPU in today's computer systems) and storage (such as DRAM), which is very often the bottleneck for computational throughput. The use of non-volatile memory also eliminates the requirement to load programs into memory during each execution. We may use, for example, the conventional flash memory device as synaptic devices [16].

Fig. 5 shows the simulation of a floating-gate synaptic transistor. As illustrated in Fig. 5(a), this special floating-gate device has separate programming terminals (negative feedback, "nfb" and positive feedback, "pfb") and synaptic gate ("sg"). This allows the synaptic gate to be always connected to the input, without the need of complicated switches to separate programming mode and read mode. The expected drawback is the reduction in coupling ratio, or charge sharing, which reduces the strength of "sg." Fig. 5(b) shows the sub-circuit for implementing this new floating-gate memory device. No extra Verilog-A code is required other than the modification of existing MOS models (a simplified version of BSIM model for use in coursework) to include Fowler-Nordheim (FN) tunneling. Even though we use FN tunneling for this example, hot carrier programming is also viable to achieve faster speed. Fig. 5(c) illustrates SPICE simulation results. The output (drain) current, which represents the weight for neural networks, can be enhanced or reduced depending on the pulse applied to the programming terminals.

Even though compact model, after calibration against experimental data, is a very accurate representation of the DC and time-dependent behavior of electronic devices, there is a limit to the number of transistor that can be included in each simulation. The maximum number of elements that can be handled with conventional SPICE is on the order of 100,000. For the typical MNIST [17] database which consists of 60,000 training sample with 28x28 handwritten digit images (784 input pixels, each with 256 levels), an estimated 228,500 synapses is required. Simulation such large circuits with 60,000 or more training cycles is beyond imagination, not to mention larger databases such as ImageNet [18]. One solution is to use FastSPICE, for which the number of elements that can be handled is on the order of 10,000,000. The other solution is to develop a dedicated neuromorphic simulation platform. NVMLearn is one example of such neuromorphic simulation platform which accounts for component level characteristics as represented by Verilog-A code, but is scalable to large deep learning problems [19]. We have tested NVMLearn on 784x1000x125x10 fully-connected neural network, with 94.3% accuracy achieved for MNIST handwritten digits recognition. Further enhancement of recognition accuracy may require convolutional neural networks, which is part of future work.

## 6. Conclusions

The Verilog-A programming language has allowed fast implementation of compact models for logic, memory devices, and beyond. Industry standard BSIM-CMG and BSIM-IMG models are developed using the surface-potential-based framework, without sacrificing computational efficiency, to model FinFET and back-gated FDSOI technologies. RRAM and floating-gate transistor compact models have also been developed for future neuromorphic applications. Continuous tuning of RRAM conductance and floating-gate current levels makes these device viable for representing the analog weights of neuromorphic circuits. To link fundamental device characteristics to system level energy, speed, and pattern recognition accuracy, a new neuromorphic simulation platform, NVMLearn, is being developed for large scale non-volatile-memory-based neural networks.

## Acknowledgements

The authors would like to express sincere gratitude to Chip Implementation Center (CIC), Hsinchu, Taiwan for providing SPICE simulation environment for RRAM simulations.